\documentstyle[11pt,newpasp,twoside,epsf]{article}
\def\edcomment#1{\iffalse\marginpar{\raggedright\sl#1\/}\else\relax\fi}
\reversemarginpar

\begin{document}
\title{Searching for X-ray Galaxy Clusters in the Zone of Avoidance} 
 \author{H. B\"ohringer, P. Schuecker, S. Komossa, J. Retzlaff, T.H. Reiprich, W. Voges}
\affil{Max-Planck-Institut f\"ur extraterr. Physik, D-85740 Garching, Germany}

\begin{abstract}
X-ray galaxy clusters are ideal tracers of the large-scale structure of
the Universe. Based on the ROSAT All-Sky Survey we are constructing a 
sample of the X-ray brightest galaxy clusters for cosmological studies.
We have already completed the  compilation of a sample of about 1000 clusters
in the sky excluding a 40 degree wide band around the galactic plane.
We have demonstrated  with the southern sample, which is the so far most complete,
that we can get reliable statistical measures of the large-scale structure of
the Universe to scales in excess of $400 h^{-1}$ Mpc. With the experience 
gained in these projects we are extending the search for X-ray galaxy 
clusters into the Zone-of-Avoidance. We have compiled a sample of 
181 good cluster candidates with a well defined completeness of about 70\%
down to a flux limit of $3\cdot 10^{-12}$ erg s$^{-1}$ in the zone with
an absorption depth of $A_v \le 2.25$. The candidates are definitely 
identified and redshifts are obtained in an ongoing optical follow-up program.
\end{abstract}

\section{Introduction}

\begin{figure}[h]  
\plotone{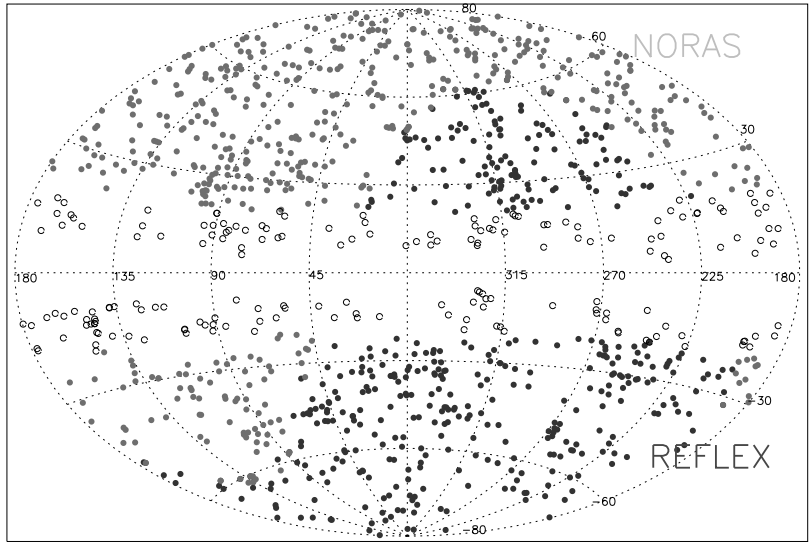}
\caption{Sky distribution of the X-ray cluster surveys based on
the RASS in galactic coordinates. The solid data points show the current
samples of the NORAS Survey (northern sky) and REFLEX Survey (southern
sky up to declination $d \le 2.5\deg$). Also shown are
the cluster candidates found in the Zone-of Avoidance $|b_{II}| \le
20\deg $ and $A_v \le 2.5$ (open circles).}
\end{figure}

Clusters of galaxies as the largest well defined building blocks of our         
Universe are ideal probes for the study of the cosmic large-scale               
structure and for the test of cosmological models (e.g. Henry et al. 1992, 
Dalton et al. 1994, Collins et al. 1995, 
Eke, Cole \& Frenk 1996, Borgani et al. 1999).
Statistical measures of the galaxy cluster population like the       
cluster mass function, the two-point-correlation function, 
and the density fluctuation power spectrum can give 
very important constraints on the characteristic measures of the matter         
density distribution throughout the Universe and its evolution as a function    
of time. For such studies the compilation of a statistically complete
cluster catalogue with a well controlled selection function is the
important first step. 

Ideally one would like to compile a cluster
catalogue where the cluster are selected by their gravitational mass.
The mass is not a direct observable, however, and one relies on 
relations of observable quantities that are closely correlated to the
cluster mass. 
While cluster catalogues have first been compiled on the basis of optical
photographic survey data (e.g. Abell 1958, Abell et al. 1989) it has been
recognized that it is very difficult to obtain a cluster mass from the
observed optical ``richness'' parameter. X-ray observations 
provide a much better means for the construction of cluster catalogues
for the study of the large-scale structure.  
The X-ray emission observed from clusters         
originates from the thermal emission of hot intracluster gas 
bound in the gravitational potential well of the clusters (e.g. Sarazin 1986).
Therefore the plasma      
emission is a certain indication of a true gravitationally bound structure.
In addition the thermal            
emission for the typical intracluster plasma                                    
temperatures of several keV has the                                             
radiative emission maximum in soft X-rays were the available X-ray              
telescopes are most effective. This makes galaxy clusters readily               
detectible out to large distances with present X-ray telescopes.  
The main advantage of the X-ray detection is, however, the fact that
the X-ray luminosity is closely correlated to the cluster mass
(Reiprich \& B\"ohringer 1999) with a dispersion of about 50\% 
in the determination of the mass for a given X-ray luminosity. 

\section{The X-ray Cluster Survey based on the ROSAT All-Sky Survey}

\begin{figure}
\plotone{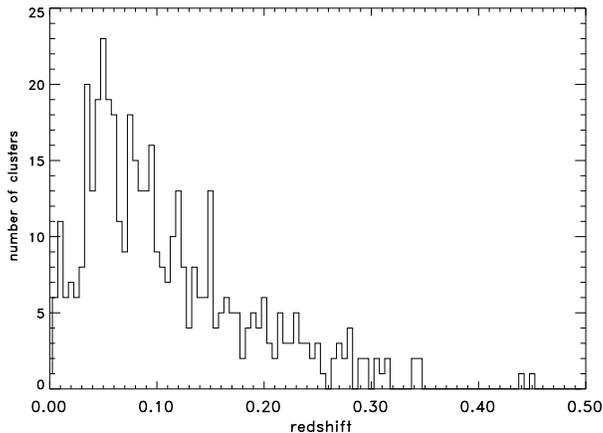}
\caption{Redshift distribution of the REFLEX clusters.} 
\end{figure}

The ROSAT All-Sky Survey (RASS), the first and up to now only All-Sky X-ray
survey conducted with an X-ray telescope (Tr\"umper 1993), 
provides the ideal starting point to compile an all-sky X-ray cluster catalogue.
To this end we are conducting a comprehensive follow-up identification 
program of the X-ray brightest galaxy clusters detected in the RASS.
Since the number of photons registered in the RASS for X-ray cluster sources
at the limit of our survey is typically about 20 - 30 photons 
this information is not sufficient to identify the nature of the X-ray
sources. Thus we supplement the X-ray detection by optical
information, which we obtain from the photographic sky surveys 
(in the north from POSS as digitized in the STScI scans and in the
south from the UK Schmidt Survey as digitized by COSMOS,
see e.g. MacGillivray et al. 1994). In addition
extensive spectroscopic measurements are necessary for the final 
identification and redshift determination.
The main part of our survey is focussed
on the region outside the band of the Milky Way
($|b_{II}| \ge 20\deg $) to avoid the regions of larger interstellar 
absorption and the more difficult optical cluster 
identification in the crowded stellar fields.

\begin{figure}
\plotone{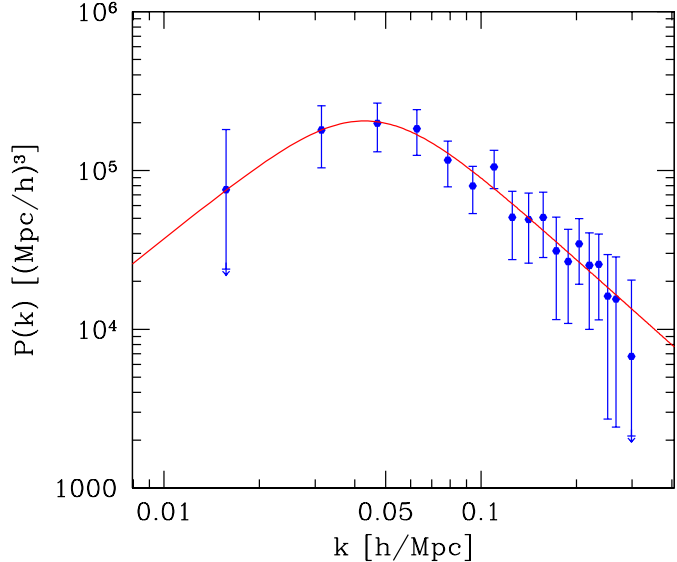}
\caption{Power spectrum of the fluctuations in the cluster density 
distribution in the REFLEX Survey (Schuecker et al. 2000).}
\end{figure}

We have recently completed the first survey steps in two major programs:
The NORAS survey in the northern sky (B\"ohringer et al. 2000a) in a
collaboration of MPE with CfA, StScI and ESO and the southern REFLEX
survey (B\"ohringer et al. 1998, 2000b, Guzzo et al. 1999) conducted as an ESO key
program with a main contribution from MPE, the Universities Milano and
Liverpool, CEA Saclay, NRL Washington, and the Royal Observatory Edinburgh 
(for results of some earlier studies related to these programs see 
also Ebeling et al. 1998, 2000, DeGrandi et al. 1999).
The sky distribution of the clusters found in the NORAS and REFLEX surveys
is shown in Fig. 1. The NORAS catalogue (B\"ohringer et al. 2000a) 
comprises so far 502 clusters. The present REFLEX sample
(B\"ohringer et al. 2000b), containing 452
clusters above a flux limit of $3\cdot 10^{-12}$ erg s$^{-1}$ cm$^{-2}$, 
is the so far best selected
and most complete X-ray survey compiled to date showing no significant
incompleteness in several tests of the selection function (see e.g. 
simulations in Schuecker et al. 2000). Therefore we apply
the REFLEX sample to a number of cosmological studies 
(e.g. Collins et al. 2000, Schuecker et al. 2000) and we will base 
the following statistics on this sample.

Fig. 2 shows the redshift distribution of the REFLEX clusters.
The median redshift is $z = 0.09 $.
The most distant cluster has a redshift of $z = 0.45$ and is the
most X-ray luminous cluster (RXCJ1347-1144) discovered to date.
The currently most important statistic describing the large-scale
structure is the density fluctuation power spectrum. Fig. 3 shows
the power spectrum for the density distribution of 188 REFLEX clusters
in a box with a length of $400 h^{-1}$ Mpc (Schuecker et al. 2000) 
showing significant structures up to scales 
of at least $400 h^{-1}$ Mpc. This is the largest volume
in the local Universe in which the structure has been studied in three
dimensions so far. The large amplitude of the power spectrum at large scales 
can only be reproduced by low density cosmological models (Schuecker et al. 
2000, Collins et al. 2000). 
In addition to the pure statistical analysis
of the large-scale structure the actual cosmography of the cluster
and matter density distribution is also studied which is particularly
important in connection with studies of the large scale flows. This
work is in progress.

\begin{figure}
\plotone{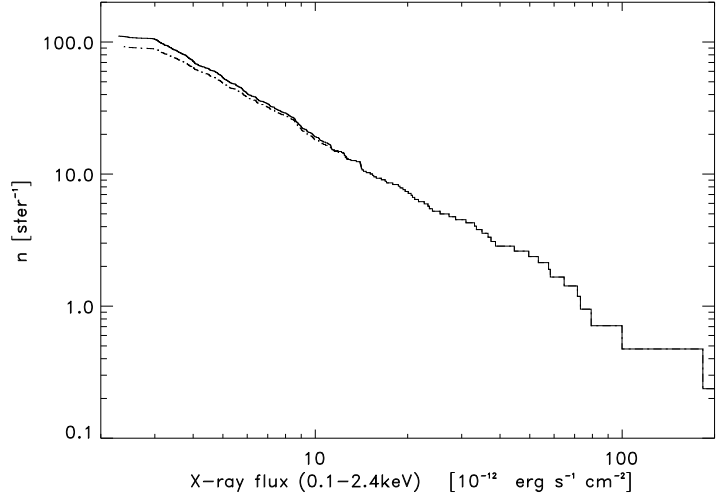}
\caption{Number counts as a function of flux ($\log$N-$\log$S-curve)
for the REFLEX sample. The solid line shows all clusters while the 
dashed line shows only the clusters displaying an extended X-ray emission.}
\end{figure}

\section{The Search for X-Ray Clusters in the Zone-of-Avoidance}

Building on the experience and the success of the REFLEX and NORAS
surveys we are extending the X-ray cluster survey also into the 
Zone-of-Avoidance (ZoA). In this region the X-ray detection is made more 
difficult by the increased interstellar absorption in the galaxy. 
Fig. 5 provides a histogram of the distribution of the interstellar
hydrogen column density in the region $|b_{II}| \le 20\deg$ (which we
loosely refer to as the ZoA in this paper even so the 
definition is somewhat different to that used in most optical surveys)
with values taken from the HI-survey of Dickey \& Lockman (1990).
We note that about 66\% of the area has a column density less than
$3 \cdot 10^{21}$ cm$^{-2}$. This corresponds to an attenuation in
the hard ROSAT band (about 0.5 - 2.0 keV, which we use for the X-ray 
cluster detection) of a factor of about 1.9
and an optical extinction of about $A_v \sim 2.25$.
Thus with a sensitivity variation of less than a factor of two we can
extend our cluster survey into about 2/3 of the ZoA area. For the start
we have restricted the systematic search to this region.

\begin{figure}
\plotone{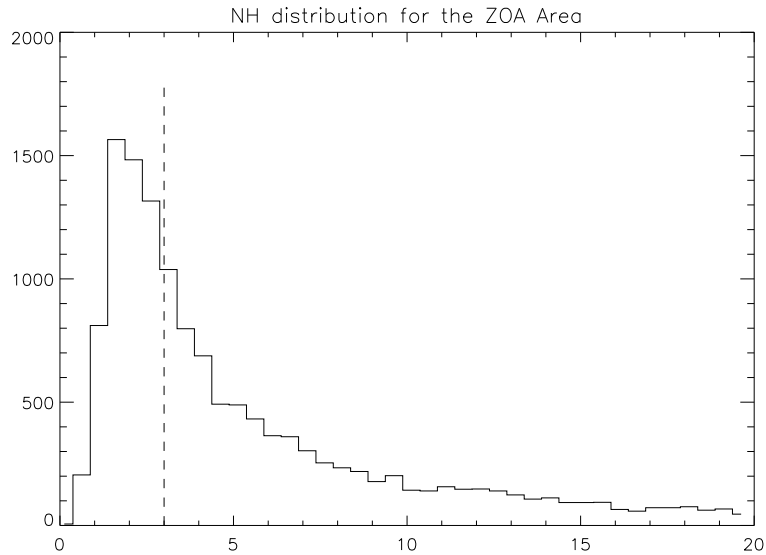}
\caption{Histogram of the distribution of the interstellar hydrogen column density
in the sky region $|b_{II}| \le 20\deg$. The vertical dashed line shows the 
current limit of our survey corresponding to an optical extinction of $A_v \sim 2.25$.}
\end{figure}

One efficient way to find cluster candidates is to look for extended 
X-ray emission. Fig. 4 shows the number counts of REFLEX clusters as a 
function of the X-ray flux limit and it also gives the fraction of clusters
that have been found as extended X-ray sources in our special analysis of 
the X-ray sources with the growth curve analysis method (GCA, see B\"ohringer
et al. 2000a). At the flux limit of $3\cdot 10^{-12}$ erg s$^{-1}$ cm$^{-2}$
81\% of the REFLEX clusters feature an X-ray extent.

Therefore we have started our search for X-ray clusters in the galactic band
among the extended X-ray sources. To this end we have reanalysed the 46404
X-ray sources detected with a likelihood $L \ge 6$ in the standard analysis
of the RASS II (the second processing of the ROSAT survey data, Voges et al. 1999). 
In total we found 358 X-ray sources with an extent probability larger than 99\% 
(based on a Kolmogorov-Smirnov method run within GCA), a flux limit $F_{lim} \ge
3\cdot 10^{-12}$ erg s$^{-1}$ cm$^{-2}$ and an interstellar absorption 
$n_H \le 3 \cdot 10^{21}$ cm$^{-2}$. From a closer inspection of these
sources in their detailed X-ray properties, their appearance on optical survey
plates, literature information, and on some follow-up observations
we can rule out 138 X-ray sources as cluster candidates. Most of these sources 
are multiple stellar X-ray sources  which are classified as extended as long as no
special care is taken to deblend these multiple sources. Other non-cluster sources 
feature a spurious extent (tests have shown that a failure rate in the extent
classification of up to 5\% has to be expected). 

\begin{figure}
\plotone{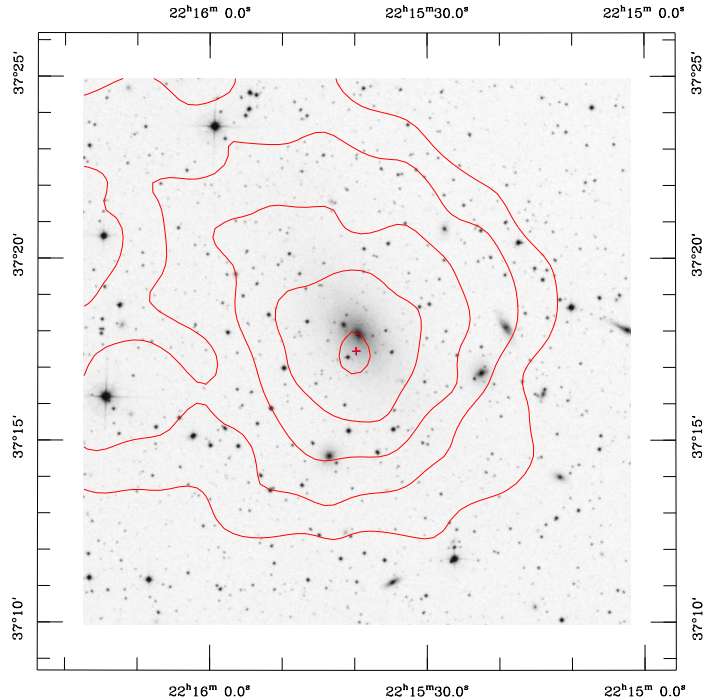}
\caption{One of the newly discovered nearby X-ray clusters at a redshift
of about $z = 0.019$. The coordinates refer to the epoch J2000. The central 
galaxy is NGC~7242 at b=$-$15.9$\deg$.}
\end{figure}

\begin{figure}
\plotone{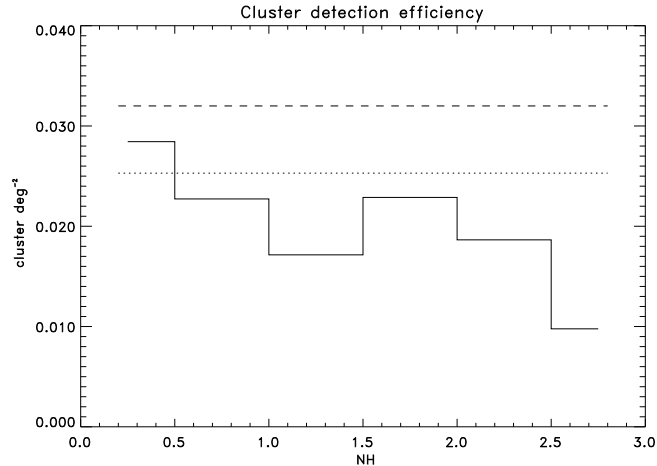}
\caption{Surface number density of the cluster candidates found in the ZoA region
(solid line). The dashed line shows the 
total REFLEX number density and the dotted line the density
of extended REFLEX clusters for comparison.}
\end{figure}

On the other hand 181 sources show cluster candidates on the optical plates. 
With the additional information that the X-ray emission of these sources
is extended the probability that these sources are clusters is very high.
For the remaining 39 X-ray sources no definite
classification could be reached and we expect that a smaller fraction of them
could be more distant clusters not easily visible on the survey plates.

The distribution of the 181 very likely cluster candidates in the sky is shown in
Fig. 1. Fig. 6 shows one of the newly found poor galaxy clusters which is 
relatively nearby at a redshift of $z \sim 0.019$. The redshift distribution
we expect for the clusters found should be similar to that shown in Fig. 2.

\section{Conclusions}

To illustrate the efficiency of the cluster detection as a function
of the interstellar absorbing column density, $n_H$, we show in Fig. 7 the 
surface density of the cluster counts in the sky as a function of $n_H$.
Also shown is the surface density of all REFLEX survey clusters 
and those which are found to be extended. In the low $n_H$ region up to
about $n_H \le 2.0 \cdot 10^{21}$ cm$^{-2}$ we almost recover the surface
density of the extended REFLEX clusters which is almost 80\% of the
total sample. Beyond $n_H \sim 2.0 \cdot 10^{21}$ cm$^{-2}$ the detection
efficiency drops significantly showing that the cluster search becomes
more difficult. In summary we estimate that we can recover at least
60 - 70\% of the clusters above a flux limit of $3 \cdot 10^{-12}$ s$^{-1}$ 
cm$^{-2}$ in 2/3 of the ZoA region which is less affected by
absorption. Definite identifications and redshifts for these cluster candidates
are currently obtained in an ongoing follow-up observation program in
the northern and southern sky.

\end{document}